# Modeling Method for the Coupling Relations of Microgrid Cyber-Physical Systems Driven by Hybrid Spatiotemporal Events

XIAOYONG BO[1,2,3], (Student Member, IEEE), XIAOYU CHEN[4], HUASHUN LI[5], YUNCHANG DONG[1,3], ZHAOYANG QU[1,3], (Member, IEEE), LEI WANG[1,3], (Member, IEEE), YANG LI[1], (Senior Member, IEEE)

[1]School of Electrical Engineering, Northeast Electric Power University, Jilin 132012, China
[2]Electrical and Information Engineering College, Jilin Agricultural Science and Technology University, Jilin 132101, China
[3]Jilin Engineering Technology Research Center of Intelligent Electric Power Big Data Processing, Jilin 132012, China
[4]State Grid Jilin Electric Power Co., Ltd. Baishan Power Supply Company, Baishan 134300, China
[5]State Grid Jilin Electric Power Co., Ltd. Jilin Power Supply Company, Jilin 132011, China

Corresponding author: Zhaoyang Qu (qzywww@neepu.edu.cn).

This work was supported in part by the Key Projects of the National Natural Science Foundation of China under Grant 51437003, and in part by the Jilin Science and Technology Development Plan Project of China under Grant 20180201092GX and 20200401097GX.

**ABSTRACT** The essence of the microgrid cyber-physical system (CPS) lies in the cyclical conversion of information flow and energy flow. Most of the existing coupling models are modeled with static networks and interface structures, in which the closed-loop data flow characteristic is not fully considered. It is difficult for these models to accurately describe spatiotemporal deduction processes, such as microgrid CPS attack identification, risk propagation, safety assessment, defense control, and cascading failure. To address this problem, a modeling method for the coupling relations of microgrid CPS driven by hybrid spatiotemporal events is proposed in the present work. First, according to the topological correlation and coupling logic of the microgrid CPS, the cyclical conversion mechanism of information flow and energy flow is analyzed, and a microgrid CPS architecture with multi-agents as the core is constructed. Next, the spatiotemporal evolution characteristic of the CPS is described by hybrid automata, and the task coordination mechanism of the multi-agent CPS terminal is designed. On this basis, a discrete-continuous correlation and terminal structure characteristic representation method of the CPS based on heterogeneous multi-groups are then proposed. Finally, four spatiotemporal events, namely state perception, network communication, intelligent decision-making, and action control, are defined. Considering the constraints of the temporal conversion of information flow and energy flow, a microgrid CPS coupling model is established, the effectiveness of which is verified by simulating false data injection attack (FDIA) scenarios.

## I. INTRODUCTION

With the advancement of the smart grid and energy Internet strategies, large amounts of electrical, sensing, and computing equipment are interconnected through grids and communication networks. Power systems with physical equipment as the core have gradually evolved into highly coupled cyber-physical systems (CPS), and are referred to as power CPS [1-2]. A power CPS is a complex network of cyber-physical integration based on the physical power grid network (which includes power primary equipment of energy flow) and cyber network (which conducts secondary control and the protection of information flow) [3]. Its main goal is to open information islands by making full use of advanced grid entities, as well as information and communication technologies related to generation, transmission, transformation, distribution, utilization, dispatching, and other links in power systems. The integration of multi-type energy systems and spatiotemporal information reflects the characteristics of holographic state perception, the ubiquitous interconnection of people and things, open platform-sharing, and internal and external business innovation [4-8].

As an important component of the smart distribution network, the microgrid is essential for smart grid construction [9-10]. The microgrid is a controllable small power system composed of micro-power sources, loads, energy storage devices, and control devices. It aims to solve the technical, market, and policy problems caused by the grid connection of large-scale and multi-type distributed power. The advantages of distributed power generation technology are exerted to the greatest extent in the economy, energy, and the environment [11-12]. The microgrid is a typical CPS; it is a comprehensive system that integrates information collection, management decision-making, control, and communication. The reliability of the physical microgrid system depends on the accurate and efficient operation of the integrated cyber system. The existence of the cyber system provides the foundation and technical guarantee for the operation of the physical system [13]. However, the dependence of the microgrid on the CPS is becoming increasingly higher, and the interaction between the information flow and energy flow of the coupled system is becoming more frequent. The CPS improves the perception, computing, communication, and control capabilities of the microgrid, while also increasing the operational safety risks of the microgrid CPS. Whether an event is a communication or proxy node failure, or a cyber network attack, it may cause physical system operation safety problems. Moreover, the failure of the physical system will cause the cyber system to lose its power supply, which will induce the cascading failure of the two networks, expand the scope of operation safety, and significantly reduce the reliability of the microgrid CPS. Therefore, based on the concept of microgrid CPS integration, in the present work, a comprehensive analysis of a microgrid is carried out, the coupling relationship is investigated with the integration mechanism, and a microgrid CPS coupled network model is established, which is of great significance for the promotion of the construction of microgrid CPS.

Existing domestic and international research on power grid CPS modeling has mainly focused on information and communication effects modeling [14-19], cyber-physical interaction modeling [20-25], and cyber-physical coupling process modeling [26-35].

Regarding information and communication effects modeling, a dynamic framework that characterizes the cyber and physical interaction between generators and loads has been proposed [14-15]. By introducing the input/output signals of physical and cyber systems, the role of information was reflected in the internal dynamic characteristics, local sensing, and execution behavior of each cyber-physical module. The reference [16] established a transient steady-state model of cyber power systems based on differential-algebraic equations, finite automata, and other mathematical tools, and, combined with the existing mathematical model of the power system, a unified power CPS model was constructed. The reference [17] analyzed the component configuration and coupling rules of an actual power network and power communication network, and established a power communication coupled network model that was consistent with the structural characteristics of the actual power CPS according to the grid topology. The reference [18-19] proposed a dual-layer multi-agent power grid CPS framework, which divides CPS clusters via cluster theory to achieve the distributed control of the power CPS by using less information.

Regarding cyber-physical interaction modeling, the reference [20] proposed a dynamic modeling method of power CPS based on a hybrid system. The actual situation of power grid CPS analysis and application was considered, and two hybrid system models, namely the finite state machine (FSM) and hybrid logic dynamics (HLD) models, were used as the power grid CPS fusion model. The reference [21] used the complete one-to-one correspondence model to describe the influence of the interaction between the power network and the cyber network. The reference [22-23] proposed methods by which to abstract the power CPS as a directed topological graph; the state quantities in the physical and cyber systems are unified as "data nodes," and the links of information processing and transmission are abstracted as an "information branch." A CPS static model was then established. The reference [24] established the degree-media weighted correlation matrix of a power CPS by defining the degree function and the electric interface, and the formal representation of a coupled CPS was realized. The reference [25] abstracted the power CPS network as a dual-layer complex network with a directed unauthorized graph, and the asymmetric balls-into-bins distribution method was used to establish a non-uniform power CPS representation model.

Regarding cyber-physical coupling process modeling, the correlation characteristic matrix method has been used to quantitatively describe the logical correlation and characteristics of the physical layer, secondary equipment layer, communication layer, and cyber layer in the power grid CPS, and a complete power grid CPS coupling model was the constructed [26-28]. The reference [29] used a matrix to describe the internal data transmission structure of the system, and then established a CPS model that considers the coupling relationship between the power and cyber networks. The reference [30] established an upward/downward communication channel model responsible for information transmission. Then, an interface model was used to associate the power grid, communication network, and cyber network. An integrated CPS model that considers multi-layer coupling was ultimately introduced. The reference [31] proposed a coupling modeling method for a power CPS based on set theory according to the logical structure analysis of a power grid CPS. The reference [32-33] put forward modeling ideas and static modeling methods of information flow, and the reference [34-35] conducted related research on the modeling and quantitative analysis of information flow.

In general, the exsiting modeling methods for power CPS are contending, and each has its own merits; some focus on information elements and consider the effects of physical processes, some focus on the description of information mapping relationships, and some focus on specific businesses. However, there currently exists no way to model a power CPS with a hybrid spatio-temporal event as the driving force, and research on the microgrid as the background of CPS modeling is rare. The essence of the microgrid CPS lies in the cyclical conversion of information flow and energy flow. Most of the existing coupling models are modeled with static networks and interface structures, while the closed-loop data flow characteristic is not fully considered. It is difficult for modeling methods that ignore dynamic behaviors (such as power flow characteristics, network attacks, etc.), coupling processes, and many other factors to accurately describe scenarios such as the spatiotemporal deduction process of the identification of an attack on the microgrid CPS network, the risk propagation mechanism, security status assessment, defense control strategies and cascading failure evolution.

Based on the preceding analysis, a coupling model of a microgrid CPS driven by a hybrid spatiotemporal event is established in the present work. The main contributions of this paper are as follows:

1) The cyclical conversion mechanism of information flow and energy flow is analyzed based on a typical microgrid structure, which improves the understanding of the closed-loop "perceive-control" process. Moreover, a microgrid CPS architecture with multi-agents is constructed, which makes the physical unit, information unit, and connection mapping more intuitive.

2) A discrete-continuous correlation and terminal structure characteristic representation method for the CPS based on heterogeneous multi-groups is proposed. Hybrid automata are used to characterize discrete-continuous correlations, and the structure of the microgrid CPS terminal is given. Multi-agents are used to characterize the structural characteristics of the CPS terminal, which can overcome the problem of traditional methods being unable to effectively characterize CPS characteristics.

3) Based on four defined spatiotemporal events, a coupling model of a microgrid CPS driven by hybrid spatiotemporal events is innovatively proposed, which compensates for the shortcomings of traditional modeling methods being unable to accurately describe the spatio-temporal deduction process of some scenes. This improves the accuracy of describing the spatiotemporal deduction process of data flow.

The remainder of this paper is organized as follows. The architecture of a microgrid CPS with multi-agents is constructed in Section II. A discrete-continuous correlation and terminal structure characteristic representation method of the CPS is proposed in Section III. The coupling model of a microgrid CPS driven by hybrid spatiotemporal events is established in Section IV. The validity of the proposed model is verified and analyzed in Section V. Finally, a summary of this paper is provided in Section VI.

## II. MULTI-AGENT MICROGRID CPS ARCHITECTURE

The typical microgrid structure based on the new situation of microgrid development is presented in Figure 1. The entire microgrid is connected to the distribution network via a grid-connected switch, which can operate independently or in parallel. First, sensors perceive the operating status of physical devices (node voltage, current, etc.), and transmit this information to agent modules for storage via data flow. Each agent module interacts with the microgrid management system via communication flow to aggregate physical-layer data. Then, the microgrid management system controls the multi-agent module of the associated unit via the Internet of Things (IoT). The agent module has a built-in decision-making unit, and the agents can be independent or communicate with each other to control global information. The agent module has self-organization and intelligent coordination capabilities, and performs optimization calculations based on the perception data, generates decision results, and sends them to the controller to generate control instructions. Finally, the physical equipment changes its operating state according to the control instructions until the entire microgrid runs smoothly [36].

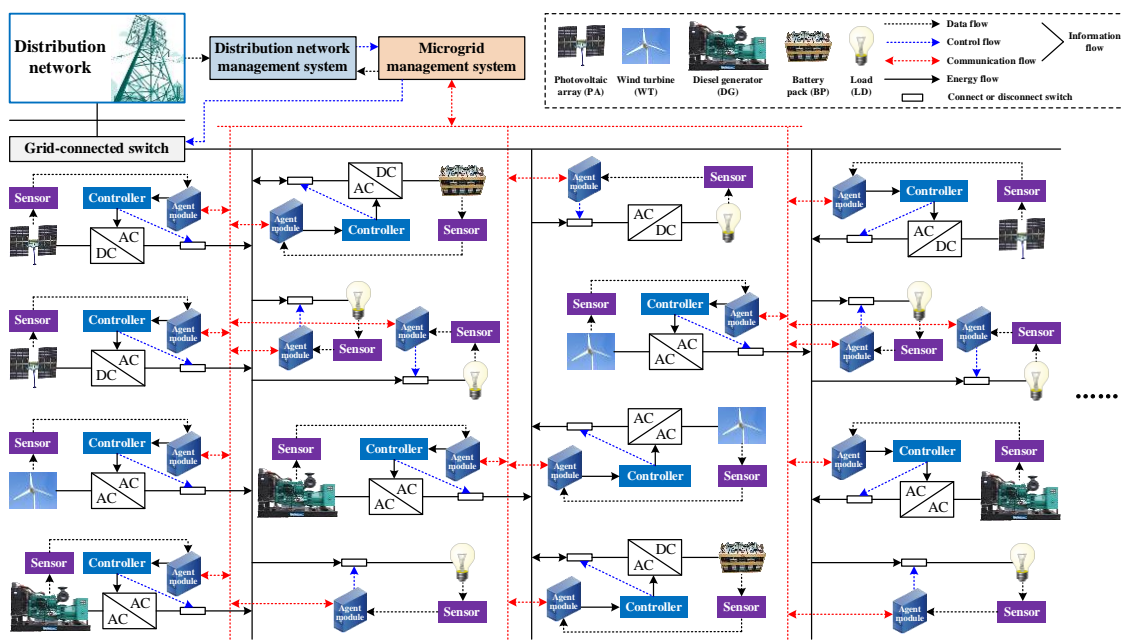

**FIGURE 1.** Schematic diagram of a typical microgrid structure.

During the entire operation of the microgrid, the conversion of information and energy is a process of "perception-control" in a continuous loop iteration. The process of sensing and acquiring the state information of the environment and physical objects by sensors can be understood as the process of transforming the energy flow of the physical system into the information flow of the cyber system. The agent module and the controller, which transmit decision-making instructions via interconnection and interoperability, are passed into the controller to accurately control the physical object. This process can be understood as the information flow of the cyber system being transformed into the energy flow of the physical system. Therefore, the entire cyclic conversion of information flow and energy flow is completed by the communication, distribution, adjustment, and coordination tasks between the multi-agent module and its sensors and controllers.

Via the preceding analysis of the mechanism of the cyclic conversion of the microgrid CPS information flow and energy flow, the main physical units, information units, and connection mapping relationships of the system were visualized, and the microgrid CPS architecture was designed, as shown in Figure 2. The architecture is based on multi-agents and consists of a two-layer structure consisting of a cyber layer and a physical layer. (1) The cyber layer consists of agent modules, terminal agent modules, and communication nodes to form the secondary control and protection of the microgrid information flow network. (2) The physical layer has different power generation units and load units, such as photovoltaic arrays, wind turbines, diesel generators, and battery packs, which constitute the energy flow network of the primary equipment of the microgrid. The two-tier structure is driven by the power IoT, including the data flow, control flow, communication flow at the cyber level, and energy flow at the physical level. During the control operation of the entire system, the information flow and energy flow are coupled with each other, forming a typical CPS.

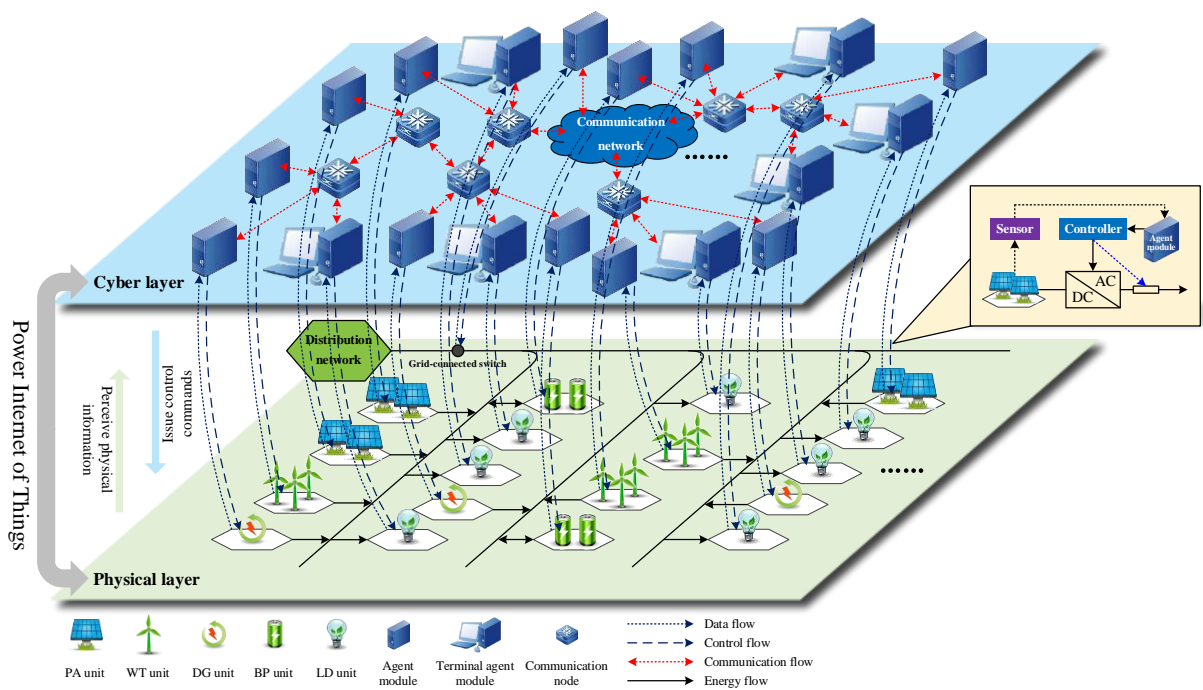

**FIGURE 2.** Architecture diagram of a multi-agent microgrid CPS.

## III. CHARACTERISTIC REPRESENTATION METHOD OF THE MICROGRID CPS

### A. DISCRETE-CONTINUOUS CORRELATION REPRESENTATION

The microgrid CPS is a close fusion of a discrete cyber space and continuous physical state. It represents a continuous physical evolution process via the action of discrete information with a spatiotemporal event-driven mechanism. Its essential feature is the interactive fusion of discrete cyber systems and continuous physical systems. The microgrid CPS has a typical spatiotemporal hybrid characteristic, which is embodied in the spatial and temporal correlation of computing and physical entities, and the sequence of events is determined by the spatial and temporal locations to ensure the correctness of the system states and behaviors. The traditional discrete-state machine cannot represent the continuous state of the system. Hybrid automata can effectively solve this problem; hybrid automata are a multivariate model that simultaneously represents the discrete and continuous states. The differential equations that represent continuous dynamic behavior are embedded into the traditional discrete-state machine model, and the automata model has the ability to characterize continuous behavior [37].

Via the use of hybrid automata, the discrete-continuous correlation of a microgrid CPS can be characterized as a seven-tuple hybrid automaton (*HA*), the expression of which is as follows:

$$HA = (D, E, C, Init, F, Inv, Act), \quad (1)$$

where $D$ represents the discrete state set of the cyber space of the microgrid CPS, $E$ represents the edge that marks the state transition, $C$ represents the continuous state set of the physical space, *Init* represents the initial state set of the system, $Init \subseteq D \times C$, $F$ represents the set of continuous dynamic equations in each discrete state $d \in D$, $D \times C \to C$, and *Inv* represents the invariant set of the mapping relationship of each discrete state $d \in D$ of the cyber space from set $D$ to set $C$. For $d \in D$, there is $Inv(d) \subset C$, and there is simultaneously a continuous state $c \in Inv(d)$. When a CPS is separated from the invariant set in the continuous action of discrete state $d$ in the cyber space, the CPS cyber space undergoes a discrete migration. Moreover, *Act* represents the mapping below $d \in D$. The interactive relationship between the continuous state variable $c$ of the CPS, its derivative $\dot{c}$, and the continuous external variable $i$ at the position $d \in D$ is expressed as $h(c, \dot{c}, i) = 0$. The solution set of this equation constitutes the activity of the microgrid CPS at $d \in D$. This equation is called the microgrid CPS state transition condition.

In the discrete-continuous correlation of a microgrid CPS based on hybrid automata, the discrete state of a hybrid automaton represents the discrete characteristics of the CPS. Each discrete state has an association invariant that keeps the hybrid automaton in this discrete state. The edge structure represents the transfer relationship between discrete states, and the defense condition (also called the transition condition) and the reset condition are used to mark the continuous state of each edge. The reset condition is that after the transition occurs, the trigger condition is reset to provide a restriction condition for the next state transition. When the defense and reset conditions are met, discrete state transitions are realized. The state of a microgrid CPS based on hybrid automata is a binary pair, which is expressed as

$$S = Q(d, c), \quad (2)$$

where $d$ represents a discrete state, $c$ represents a continuous variable, $c \in C$, and $c$ satisfies $Inv(c)$.

The two transition states of the microgrid CPS are supposedly based on hybrid automata $S_m = Q(d_m, c_m)$ and $S_n = Q(d_n, c_n)$, respectively. If there is an $f \in F : d_m \to d_n$ state transition, and the values of $c_m, c_n$ satisfy the defense and reset conditions on *f*, this transition $(Q_m, Q_n)$ is called a jump, and $Q_n$ is the jump successor of $Q_m$. This process can characterize the interaction between the cyber and physical space of the microgrid CPS.

### B. CHARACTERIZATION OF THE CPS TERMINAL STRUCTURE

During the operation of the microgrid CPS, the cyber space agent module exhibits certain self-organization and intelligent coordination capabilities, which are the basic characteristics of a multi-agent system. A single agent module makes a real-time and accurate response to the surrounding environmental conditions. It can handle conflicts with other agent modules, or coordinate with other agent modules to handle conflicts, plan behaviors, and finally make decisions. The agent module perceives the environment through sensors, and affects the environment by executing the actions of the controller to achieve its goals.

The microgrid CPS system architecture includes multiple information and physical primitives, such as agent modules, sensors, and controllers. The intelligent fusion of information and physical primitives forms a decision-making organization. In this paper, the implementation of decision-making agencies, the perception of external information, and the feedback of the implementation results are defined as the CPS terminal of the microgrid. The microgrid CPS can quickly respond to external requests or control objects by sensing external environmental information, collecting physical object status information, and combining information and physical fusion to obtain decisions. Simultaneously, it can coordinate and handle conflicts with other CPS terminals, plan behaviors, and complete tasks together to accurately control physical objects. Based on the constructed microgrid CPS system architecture, the CPS terminal structure is presented in Figure 3.

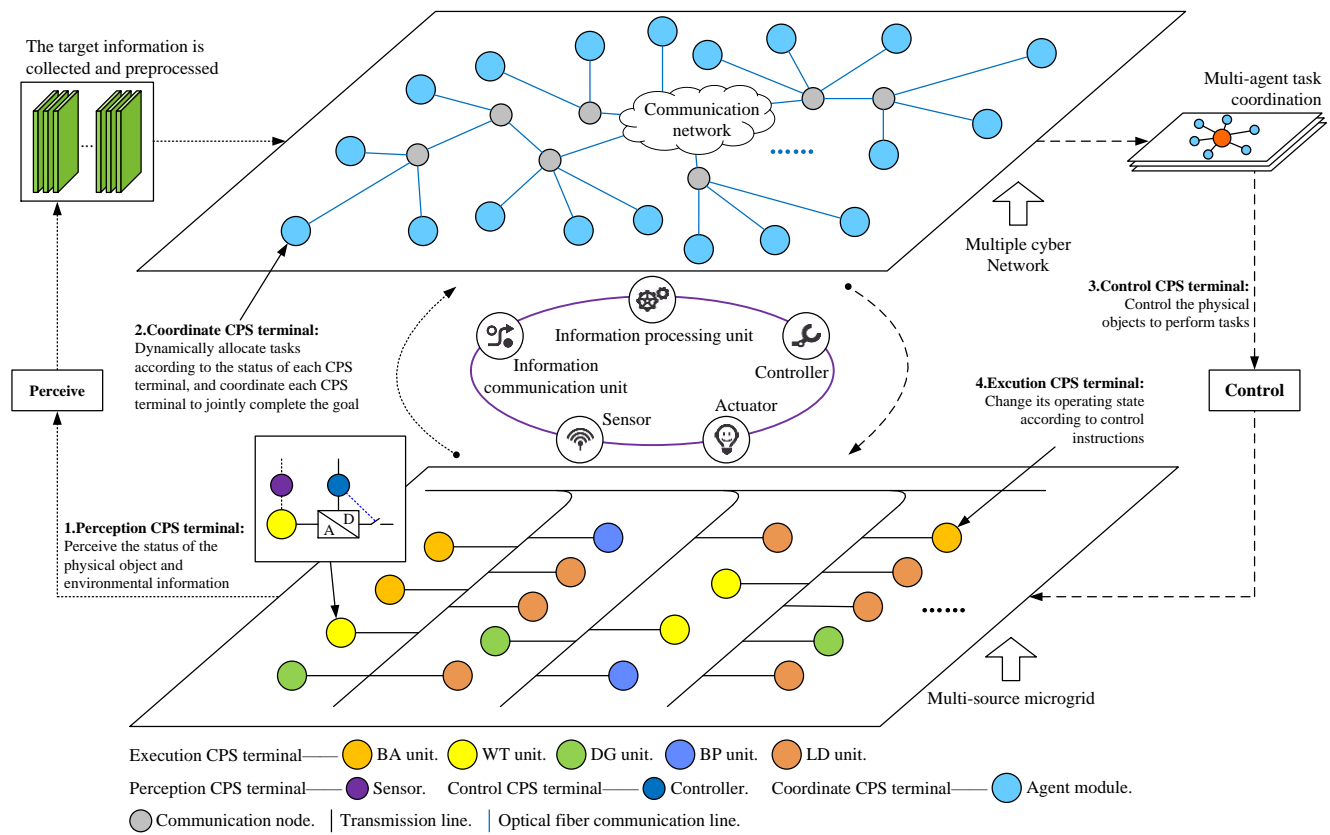

**FIGURE 3.** Structural diagram of the microgrid CPS terminal.

According to different functions, the microgrid CPS terminal can be categorized as a perception CPS terminal, control CPS terminal, coordination CPS terminal, or execution CPS terminal. Among them, the perception, control, and execution CPS terminals use mutual communication and status feedback for information or energy conversion, or control the physical object to perform tasks, perceive the status of the physical object and environmental information. The coordination CPS terminal dynamically allocates tasks according to the status of each CPS terminal, and coordinates each CPS terminal to jointly complete the goal. There are differences in the completion of different tasks by the CPS terminal, which are manifested in differences in size, location, and function. In the structure of the microgrid CPS terminal, the target information is collected and preprocessed by the perception CPS terminal (the sensor, etc.), and the coordination CPS terminal of the multiple cyber networks (the information communication unit, information processing unit, etc.) performs multi-agent task coordination according to the state data of the multi-source microgrid. The control CPS terminal (the controller, etc.) controls the execution CPS terminal (the actuator, etc.) to change its operating state according to the task coordination decision result. The entire structure of the microgrid CPS terminal is a coupled cyclic process based on the conversion of information flow and energy flow driven by spatiotemporal events.

The CPS terminal mainly completes information and energy conversion tasks during system operation, and is an indispensable component of the CPS. In this paper, by combining the functional characteristics of the multi-agents and the CPS terminal, the structural characteristics of the CPS terminal of the microgrid can be characterized as a ten-tuple $DC$, the expression of which is as follows:

$$DC = (D_{ID}, Fu, RT, OT, A, St, Ac, VI, EI, R_{FB}), \quad (3)$$

where $D_{ID}$ represents the unique identification of the different functional terminals of the microgrid CPS, $Fu$ describes the function of the single CPS terminal, $RT$ represents the real-time state set of the single CPS terminal in the process of completing tasks and real-time interaction, $OT$ represents the task set of the single CPS terminal, $A$ represents the set of goals that a single CPS terminal needs to achieve, $St$ represents the set of strategies that a single CPS terminal should take to complete the task, $Ac$ represents the set of actions performed to complete the task, $VI$ represents the set of various information required by the single CPS terminal to complete the task, $EI$ represents the information set obtained from the external environment, and $R_{FB}$ represents the feedback information set of the task execution result.

## IV. COUPLING MODELING OF THE MICROGRID CPS

The microgrid is based on small primary power systems

(including micro power supplies, energy storage devices, loads, etc.), and integrates numerous sensors, controllers, actuators, communication equipment, and decision-making units. It is a typical multi-layer heterogeneous complex system in which the coupling relationship between layers is closer, and the interaction between information and physics is more frequent. Due to this complexity and its multi-scale characteristics in time and space, the entire control process of the microgrid CPS is susceptible to many factors, such as the communication channel capacity, network connectivity, and reliability. Overall, the microgrid CPS can be divided into a static model and dynamic model. In the present work, the microgrid CPS is modeled from the two aspects of a static coupled network model and a dynamic state transition model. The model framework is presented in Figure 4.

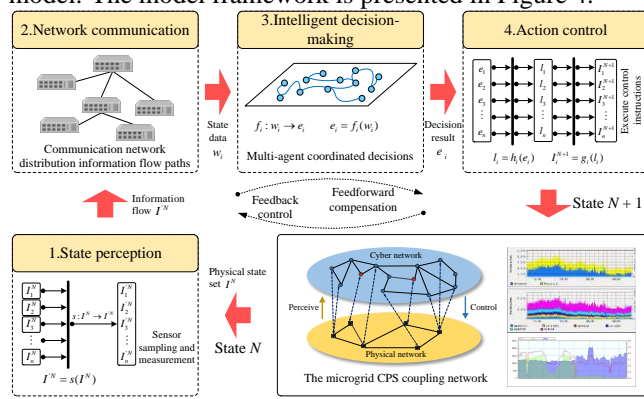

**FIGURE 4.** Framework diagram of the microgrid CPS coupling model.

The initial state of the microgrid CPS coupling network is $N$, and the physical state set can be expressed as $I^N = [I_1^N, I_2^N, ..., I_n^N]$. First, the coupling network enters the state perception process, and the information flow $I'^N$ is obtained via sensor sampling and measurement. The function $s$ realizes the conversion of physical state information to digital signals, which represents the conversion of energy flow to information flow. Next, the information flow is routed through network communication. The information flow passes through multiple routing jumps to provide state data $w_i$ for intelligent decision-making. Then, the intelligent decision-making process makes multi-agent coordinated decisions based on the state data set $[w_1, w_2, ..., w_n]$, and obtains the decision result $e_i$. The function $f$ represents the decision function integrated into the corresponding multi-agent collaboration algorithm. Finally, the decision result $e_i$ is used as the input of the control instruction conversion function $h_i$ to generate control instructions, and the control commands are applied to the physical power grid by the function $g_i$. The microgrid CPS coupling network then enters the $N + 1$ state, which represents the conversion of information flow to energy flow. From the perspective of the control system, this can be regarded as a comprehensive process based on the combination of CPS spatiotemporal event-driven feedforward compensation and feedback control. Although the occurrence of the event is separated in time and space, the spatiotemporal information of the event is synchronized during the process of collection, transmission, and processing.

### A. STATIC COUPLING NETWORK MODELING
The microgrid CPS is a typical dependent cyber-physical network. The physical network relies on the dispatch control of the cyber network, and the cyber network relies on the energy supply of the physical network. The entire coupling network is a fused cyber-physical cyclic network driven by a spatiotemporal event, and is part of a one-to-one coupling network. Via the application of complex network theory, physical and cyber network models can be abstracted into topological diagrams.

#### 1) ABSTRACT TOPOLOGICAL MODEL OF A PHYSICAL NETWORK
Physical microgrid units, such as photovoltaic arrays, wind turbines, diesel generators, and battery packs, can be abstracted as power a node $V_{mp}$. Power users, such as power transmission lines, can be abstracted as a load node $V_{ml}$, point set $V_m = (V_{mp}, V_{ml}, ...)$, and edge set $E_m$. The abstract topology of the physical network can be represented as a graph $G_m = (V_m, E_m)$ consisting of a point set and an edge set.

#### 2) ABSTRACT TOPOLOGICAL MODEL OF A CYBER NETWORK
Similar to the construction of an abstract topological model of a physical network, microgrid information units, such as agent modules and terminal agent modules, can be abstracted as a core node $V_{ic}$. Additionally, routers and switches, such as optical fiber transmission lines, can be abstracted as a transmission node $V_{it}$, point set $V_i = (V_{ic}, V_{it}, ...)$, and edge set $E_i$. The abstract topological graph of the cyber network can be expressed as a graph $G_i = (V_i, E_i)$ consisting of a point set and edge set.

#### 3) STATIC COUPLING NETWORK MODEL OF A MICROGRID CPS
Via the abstract topological models of the physical and cyber networks, the actual microgrid CPS component configuration rules are combined to construct a static coupled network model of the microgrid CPS, as follows.

(1) An abstract topological model is established for the physical unit of the microgrid.

(2) The number of core nodes in the cyber network $n$ is determined, which is equal to the number of physical units of the microgrid, and an abstract topology model of the cyber network is established.

(3) The physical and cyber networks are mapped and connected, i.e., the physical node $p$ corresponds to the cyber node $i$.

Based on the preceding content, suppose that there are no multiple edges in the coupling network, and that the coupled cyber-physical network is abstracted as a graph $G_{ip} = (V_{ip}, E_{ip})$ composed of a point set $V_{ip}$ and an edge set $E_{ip}$. The topological structure of $G_{ip}$ is represented by an adjacency matrix $M$:

$$M = \begin{bmatrix} M_{ii} & M_{ip} \\ M_{pi} & M_{pp} \end{bmatrix}, \quad (4)$$

where $i$ represents a cyber network node and $p$ represents a physical network node. Each sub-block in $M$ is the adjacency matrix between corresponding node sets.

### B. DYNAMIC STATE TRANSITION MODEL

The "perception-control" process of the microgrid is a closed-loop control process based on the cyclic conversion of information flow and energy flow driven by spatiotemporal events. It follows the "perception-communication-decision-control" temporal conversion constraint. A certain operating cycle is taken as an example, four spatiotemporal events are defined, and a dynamic model of the state transition of each event is established.

#### 1) MODEL OF A STATE PERCEPTION SPATIOTEMPORAL EVENT

The perception CPS terminal samples and measures the physical units of the microgrid in the $N$ state. In this paper, this spatiotemporal event is defined as a state perception spatiotemporal event. The initial physical state set is $I^N = [I_1^N, I_2^N, ..., I_n^N]$, and the energy flow is realized from time and space to information flow via the $s$ function. The information flow is obtained after sampling, and the mathematical expression is as follows.

$$I'^N = s(I^N) \quad (5)$$

#### 2) MODEL OF A NETWORK COMMUNICATION SPATIOTEMPORAL EVENT

For the information flow $I_i'^N$ sampled by physical node $i$, the information jump path is determined by the routing algorithm, and the network communication sends the information flow $I_i'^N$ to the coordinate CPS terminal to provide the required data $w_i$ for an intelligent decision-making spatiotemporal event. In this paper, this spatiotemporal event is defined as a the network communication spatiotemporal event, the mathematical expression of which is

$$w_i = r_i(I_i'^N), \quad (6)$$

where $r_i$ represents the jump path function of the sampling information flow of physical node $i$ during network communication.

#### 3) MODEL OF AN INTELLIGENT DECISION SPATIOTEMPORAL EVENT

As shown in Figure 5, each core node $[V_{i1}, V_{i2}, ..., V_{in}]$ of the cyber space receives the state data set $[w_1, w_2, ..., w_n]$, and uses the microgrid multi-agent decision-making methods, such as reinforcement learning and the Monte Carlo $Q$ function, to obtain the decision result $e_i$.

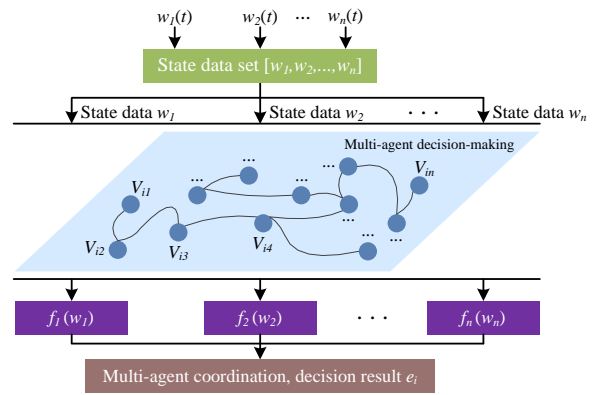

**FIGURE 5.** Intelligent decision-making process of a microgrid coordinate CPS terminal.

This spatiotemporal event is defined as an intelligent decision-making spatiotemporal event, the mathematical expression of which is

$$e_i = f_i(w_i), \quad (7)$$

where $f_i$ represents the decision function integrated into the corresponding multi-agent collaboration algorithm. It can realize not only the self-organizing decision-making of the core node, but also coordinated decision-making with other core nodes.

#### 4) MODEL OF A ACTION CONTROL SPATIOTEMPORAL EVENT

The decision result of the core node must act on the corresponding physical unit via action control to drive the state change of the physical space. This spatiotemporal event is defined as an action control spatiotemporal event.

First, the decision result $e_i$ is input into the control instruction conversion function $h_i$ to generate the control instruction $l_i$, the mathematical expression of which is as follows.

$$l_i = h_i(e_i) \quad (8)$$

Then, the control command is applied to the physical power grid by the function $g_i$, and the micro-grid CPS coupling network enters the $N + 1$ state, the mathematical expression of which is as follows.

$$I_i^{N+1} = g_i(l_i) \quad (9)$$

#### 5) MODEL OF A STATE TRANSITION CONTINUOUS SPATIOTEMPORAL EVENT-DRIVEN

The correct operation of the microgrid CPS is closely related to the accurate execution of state perception, network communication, intelligent decision, and action control. By synthesizing the spatiotemporal event models of each link described previously, a continuous spatiotemporal event-driven model of state transition can be obtained, the mathematical expression of which is

$$I_i^{'N} = g_i(h_i(f_i(r_i(s(I^N))))) = \delta_i(I^N), \quad (10)$$

where $\delta_i$ represents the continuous driving function of state transition from state $N$ to state $N + 1$ in one operating cycle of the microgrid CPS.

## V. EXAMPLE ANALYSIS

This section reports the construction of a simulation environment for the microgrid CPS, and the false data injection attack (FDIA) scenario is then considered as an example. Based on the proposed model, the spatiotemporal deduction process of the attack data stream in the microgrid CPS is accurately described, and the model effectiveness and rationality are verified.

Figure 6 presents the structural diagram of the simulated microgrid simulation. The physical model includes 3 distributed generations, 3 loads, and transmission lines. Each physical unit comprises an integrated sensor and controller. The cyber model includes 3 routers, 6 agent nodes, and network transmission lines. Information units are connected by routers. Each physical unit has a one-to-one correspondence with the local agent module. The agent module has the function of collecting local measurements, and can exchange information with other adjacent agent modules via network communication.

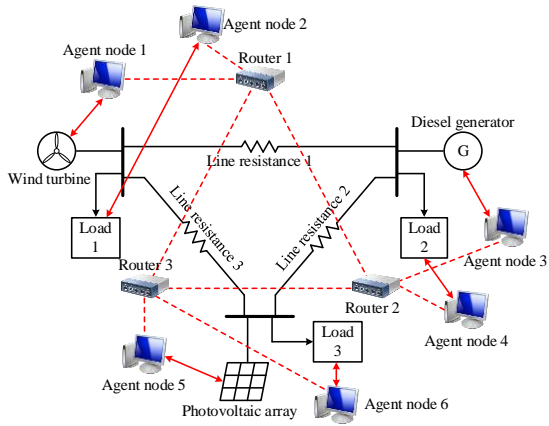

**FIGURE 6.** Structural diagram of the simulated microgrid system.

### A. SIMULATION ENVIRONMENT CONSTRUCTION

The simulation environment was constructed by the RT-LAB real-time simulation system and OPNET network simulation tool. The entire simulation environment was composed of four main components, namely the RT-LAB host, OPAL-RT OP4510 target machine, OPNET host, and intelligent decision workstation, and each component was connected by an Ethernet switch. The simulation environment is shown in Figure 7, and the basic design concept is as follows. In each operating cycle of the microgrid CPS, the simulation data of the physical space of the microgrid is sent to the simulated cyber space system via Ethernet, and the real-time data drive the cyber space calculation process toward intelligent decision-making. Then, the simulated cyber space system returns the decision instruction to the simulated physical space system, and finally realizes the change of the operating state of the physical system.

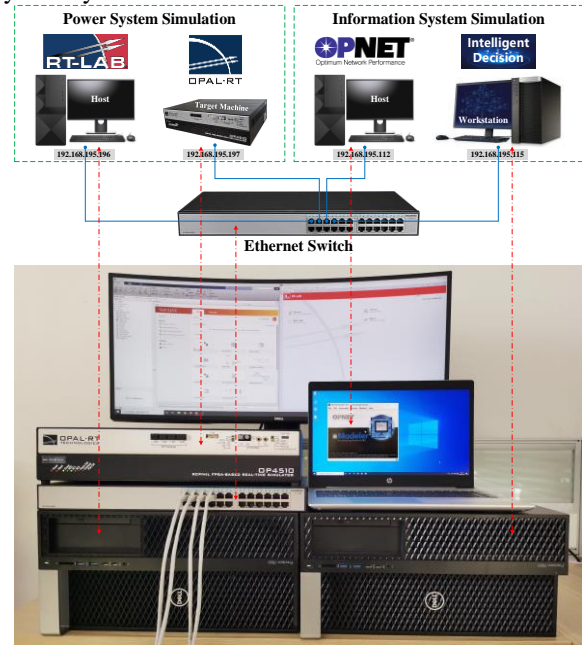

**FIGURE 7.** Simulation environment of microgrid CPS.

The RT-LAB host was installed with the simulation software RT-LAB, and the microgrid physical model was constructed by running MATLAB/Simulink on the software. The model was finally converted into the C language code and downloaded to the OPAL-RT OP4510 target machine. The multi-core and high-frequency advantages of the OPAL-RT OP4510 target machine were used to realize the real-time simulation of the physical system of the microgrid.

The OPAL-RT OP4510 target machine exchanged data with the RT-LAB host, OPNET host, and intelligent decision workstation via Ethernet. The N nodes on the OP4510 target machine represent N physical units. The data measurement model on each unit was responsible for collecting the voltage, current, and other simulation data. The data-sending model was responsible for sending the simulation data to the intelligent decision workstation via the Socket communication port, and the data-receiving model was responsible for receiving the control commands sent by the intelligent decision workstation via the Socket communication port that called the OPNET information node.

The OPNET host simulated the cyber network of the microgrid CPS. The network contained N agent nodes corresponding to communication devices. The data-mapping module on each agent node was responsible for sending the control instructions generated by the intelligent decision-making workstation to the OPAL-RT OP4510 target machine. It was also responsible for monitoring the communication status of the microgrid CPS cyber network.

The intelligent decision-making workstation was responsible for conducting both single-point and multi-agent decision-making via the received measurement data, and for generating control instructions corresponding to the physical unit. By calling the OPNET information node sending module, the action control was applied to the physical unit, which had an impact on the running state of the physical unit.

### B. SIMULATION EXPERIMENT AND RESULT ANALYSIS

According to the established static model of the coupling network, the abstract topological model of the coupling network of the simulated microgrid system shown in Figure 6 can be expressed as a matrix.

$$M = \begin{bmatrix} 1 & 0 & 0 & 0 & 0 & 0 & 1 & 0 & 0 & 1 & 0 & 0 & 0 & 0 \\ 0 & 1 & 0 & 0 & 0 & 0 & 1 & 0 & 0 & 0 & 1 & 0 & 0 & 0 \\ 0 & 0 & 1 & 0 & 0 & 0 & 0 & 1 & 0 & 0 & 0 & 1 & 0 & 0 \\ 0 & 0 & 0 & 1 & 0 & 0 & 0 & 1 & 0 & 0 & 0 & 0 & 1 & 0 \\ 0 & 0 & 0 & 0 & 1 & 0 & 0 & 1 & 0 & 0 & 0 & 0 & 0 & 1 \\ 0 & 0 & 0 & 0 & 0 & 1 & 0 & 0 & 1 & 0 & 0 & 0 & 0 & 1 \\ 1 & 1 & 0 & 0 & 0 & 0 & 1 & 1 & 1 & 0 & 0 & 0 & 0 & 0 \\ 0 & 0 & 1 & 1 & 0 & 0 & 1 & 1 & 1 & 0 & 0 & 0 & 0 & 0 \\ 0 & 0 & 0 & 0 & 1 & 1 & 1 & 1 & 1 & 0 & 0 & 0 & 0 & 0 \\ 1 & 0 & 0 & 0 & 0 & 0 & 0 & 0 & 0 & 1 & 1 & 1 & 1 & 1 \\ 0 & 1 & 0 & 0 & 0 & 0 & 0 & 0 & 0 & 1 & 1 & 1 & 1 & 1 \\ 0 & 0 & 1 & 0 & 0 & 0 & 0 & 0 & 0 & 1 & 1 & 1 & 1 & 1 \\ 0 & 0 & 0 & 1 & 0 & 0 & 0 & 0 & 0 & 1 & 1 & 1 & 1 & 1 \\ 0 & 0 & 0 & 0 & 1 & 0 & 0 & 0 & 0 & 1 & 1 & 1 & 1 & 1 \\ 0 & 0 & 0 & 0 & 0 & 1 & 0 & 0 & 0 & 1 & 1 & 1 & 1 & 1 \end{bmatrix} \quad (11)$$

The FDIA was selected as a typical network attack, and, based on the proposed dynamic state transition model, the spatiotemporal deduction process of the attack data stream in the microgrid CPS was accurately described.

*Scenario 1: Measurement data injection attack.* According to the cyclic conversion process of information flow and energy flow, a measurement data injection attack acts on state perception spatiotemporal events in the "perception-communication-decision-control" temporal conversion process of the microgrid CPS. The spatiotemporal deduction process of the attack data flow is as follows. An attacker invades the system through the potential intrusion position of the cyber network, obtains the control rights of the uplink communication channel/RTU sensor of the agent node, and conducts a measurement data injection attack. This attack is defined as tampering with the measured system parameters or operating information, resulting in the mismatch between the data information received by the agent node via network communication and the actual operating state of the system, so that the agent node makes incorrect decisions based on the erroneous information.

Based on the constructed microgrid CPS simulation environment described in Section 4.1, agent node 4 was selected. Via sensor perception and network communication, agent node 4 can collect the A, B, and C voltage (current) phase angles and phase magnitudes of load 2, as well as the positive-, negative-, and zero-sequence voltage (current) phase angles, phase magnitudes, etc. The positive-, negative-, and zero-sequence current phase angles were taken as an example. Table 1 presents the changes of the temporal positive-, negative-, and zero-sequence current phase angles of load 2.

**TABLE 1.** Positive-, negative-, and zero-sequence current phase angles.

| Time number | Positive-sequence current phase angle (°) | Negative-sequence current phase angle (°) | Zero-sequence current phase angle (°) |
|---|---|---|---|
| 1  | 116.29897 | 0         | 0         |
| 2  | 116.08698 | 0         | 0         |
| 3  | 118.21265 | 0         | 0         |
| 4  | 118.23557 | 0         | 0         |
| 5  | 117.86888 | 0         | 0         |
| 6  | 117.84023 | 0         | 0         |
| 7  | 117.71991 | 0         | 0         |
| 8  | 117.66834 | 0         | 0         |
| 9  | 117.53083 | 0         | 0         |
| 10 | 114.92388 | 0         | 0         |
| 11 | 109.56672 | 165.00611 | -58.24689 |
| 12 | 117.41051 | 171.64097 | -49.97911 |
| 13 | 118.05795 | 172.59208 | -50.25413 |
| 14 | 118.31006 | 173.92134 | -52.28240 |
| 15 | 117.61105 | 171.90453 | -53.77782 |
| 16 | 117.55948 | 171.32011 | -52.27094 |
| 17 | 117.44489 | 170.54089 | -51.25680 |
| 18 | 117.03236 | 166.54737 | -56.04673 |
| 19 | 116.88912 | 169.71583 | -50.52342 |
| 20 | 116.76307 | 168.32354 | -54.87790 |

First, $I_{up}^{Pos}$, $I_{up}^{Neg}$, and $I_{up}^{Zero}$ are respectively defined as the positive-, negative-, and zero-sequence current phase angles of the measurement information. If there is no attack, then $I_{up}^{Pos} > 0$, $I_{up}^{Neg} = I_{up}^{Zero} = 0$. If the uplink communication channel/RTU sensor corresponding to agent node 4 suffers a measurement data injection attack, the negative- and zero-sequence current phase angles of load 2 are tampered with, $I_{up}^{Neg} > 0$ and $I_{up}^{Zero} < 0$, respectively, and $I_{up}^{Neg}$ and $I_{up}^{Zero}$ are not equal to 0. According to the proposed state perception spatiotemporal event model $I^{'N} = s(I^{N})$, $I_{up}^{Pos}$, $I_{up}^{Neg}$, and $I_{up}^{Zero}$, the error measurement information is converted from time and space to the information flow via the $s$ function, and the information flow $I_{up}^{'Pos}$, $I_{up}^{'Neg}$, $I_{up}^{'Zero}$ is obtained after sampling. As shown in Figure 8, at time number 1-10, the positive-sequence current phase angle is positive, and the energy flow is normally transmitted. Starting at time 11, the attacker tampered with the negative- and zero-sequence current phase angles of load 2 via the measurement data injection attack, causing them to have positive and negative values, respectively. In this case, the positive-, negative-, and zero-sequence current phase angles were collected to agent node 4 via network communication. After analysis and decision-making, agent node 4 believed that the load 2 node was faulty, but, in fact, the load 2 node was operating normally. Because the attacker tampered with the original normal data via an FDIA, agent node 4 may eventually make an incorrect wrong decision.

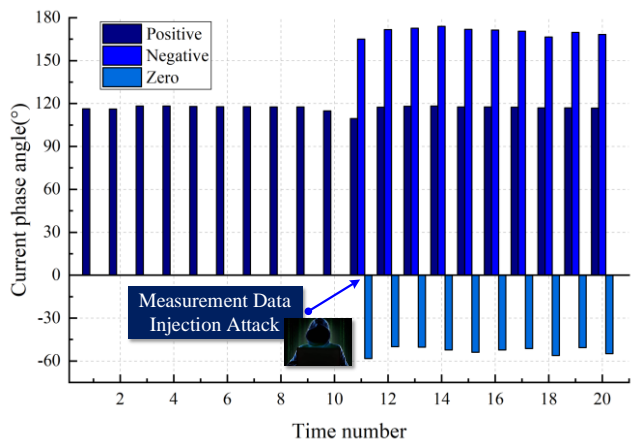

FIGURE 8. Measurement data injection attack tampering with negative- and zero-sequence current phase angles.

Figure 9 presents the three-dimensional surface comparison diagram of the topological voltage phase angles of the microgrid simulation system before and after the measurement data injection attack. It can be seen from the figure that the three-dimensional surface diagram of the voltage phase angle were significantly different before and after the measurement data injection attack.

*Scenario 2: Remote control command injection attack.* According to the cyclic conversion process of information flow and energy flow, the remote control command injection attack acts on the action control spatiotemporal event in the "perception-communication-decision-control" temporal conversion process of the microgrid CPS. The spatiotemporal deduction process of the attack data flow is as follows. An attacker intrudes into the system via the potential intrusion position of the cyber network, obtains the control rights of the downlink communication channel/RTU actuator of the agent node, and conducts a remote control command injection attack. This attack is defined as tampering with intelligent decision control commands (such as the generator output, circuit breaker opening and closing, etc.), which causes physical equipment to execute false control instructions and ultimately, directly affects the operation of the power grid.

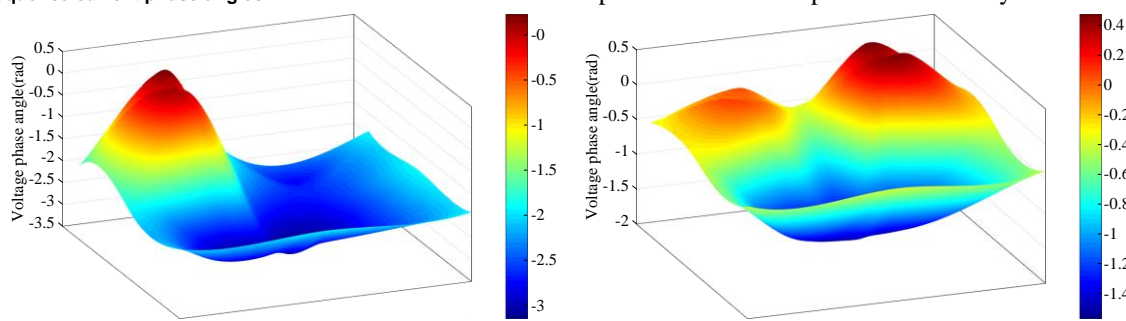

FIGURE 9. Comparison of the three-dimensional surface diagram of the voltage phase angle before and after the measurement data injection attack.

Based on the constructed microgrid CPS simulation environment described in Section 4.1, agent node 4 was selected, which obtains the decision result and generates control instructions via self-organizing or multi-agent decision-making. The opening and closing instructions of the circuit breaker were taken as an example, and Table 2 reports the changes of the voltage and current phase magnitudes with time.

First, $e_4^{down}$ and $l_4^{down}$ are defined as the decision result and control command of agent node 4, respectively. If there is no attack, then $e_4^{down} = E_4^{down}$, where $E_4^{down}$ is the decision result value, and $l_4^{down} = L_4^{down}$, where $L_4^{down}$ is the control command value. If the downlink communication channel/RTU actuator corresponding to agent node 4 suffers a remote control command injection attack, $l_4^{down} = L_4^{down} + c$, where $c$ is the tampering factor, which means that the attacker has realized the opening or closing operation of the circuit breaker via the tampering factor.

TABLE 2. Changes of the voltage and current phase magnitudes with time.

| Time number | Voltage phase magnitude (V) | Current phase magnitude (A) |
|---|---|---|
| 511 | 401.0963 | 0.8068 |
| 512 | 401.0963 | 0.8062 |
| 513 | 401.0963 | 0.8068 |
| 514 | 401.0963 | 0.8062 |
| 515 | 415.3805 | 1.4199 |
| 516 | 391.5229 | 1.5287 |
| 517 | 392.9665 | 1.5420 |
| 518 | 393.7263 | 1.5437 |
| 519 | 395.0940 | 1.5470 |
| 520 | 395.2459 | 1.5503 |
| 521 | 395.7778 | 1.5553 |
| 522 | 395.7778 | 1.5559 |
| 523 | 395.9297 | 1.5642 |
| 524 | 350.0380 | 3.7654 |
| 525 | 426.3216 | 0.8245 |
| 526 | 426.5495 | 0.8395 |
| 527 | 396.3856 | 1.6441 |
| 528 | 396.0817 | 1.5947 |
| 529 | 396.0817 | 1.5858 |
| 530 | 396.0817 | 1.5842 |
| 531 | 396.0057 | 1.5808 |
| 532 | 395.9297 | 1.5814 |
| 533 | 395.9297 | 1.5814 |
| 534 | 418.4957 | 1.0593 |
| 535 | 405.8831 | 1.0853 |
| 536 | 403.6037 | 1.0953 |
| 537 | 400.5645 | 1.1075 |
| 538 | 397.8292 | 1.1242 |
| 539 | 396.9934 | 1.1347 |
| 540 | 398.4371 | 1.1747 |

According to the proposed action control spatiotemporal event model $l_i = h_i(e_i)$ and $I_i^{N+1} = g_i(l_i)$, $e_4^{down}$, $l_4^{down}$ through the control command conversion function $h_i$ and the control command execution function $g_i$, and the microgrid CPS coupling network is made to enter the $N+1$ state, which is the attacker's intended state. As shown in Figure 10, at time number 1-510, the voltage and current phase magnitudes fluctuated normally. Starting at the time 511, the attacker tampered with the intelligent decision-making control command of agent node 4 via the remote trip command injection attack, causing it to execute the circuit breaker opening action, which caused load-shedding. In this case, load 2 received the tampered control command, and the physical device executed the opening action. The instant load-shedding caused grid instability, the voltage and current fluctuated greatly compared with the previous continuous fluctuations, and the power flow was redistributed. In severe cases, the instantaneous opening of the circuit breaker will cause the circuit to be overloaded and tripped, which will cause a large-scale blackout in the area.

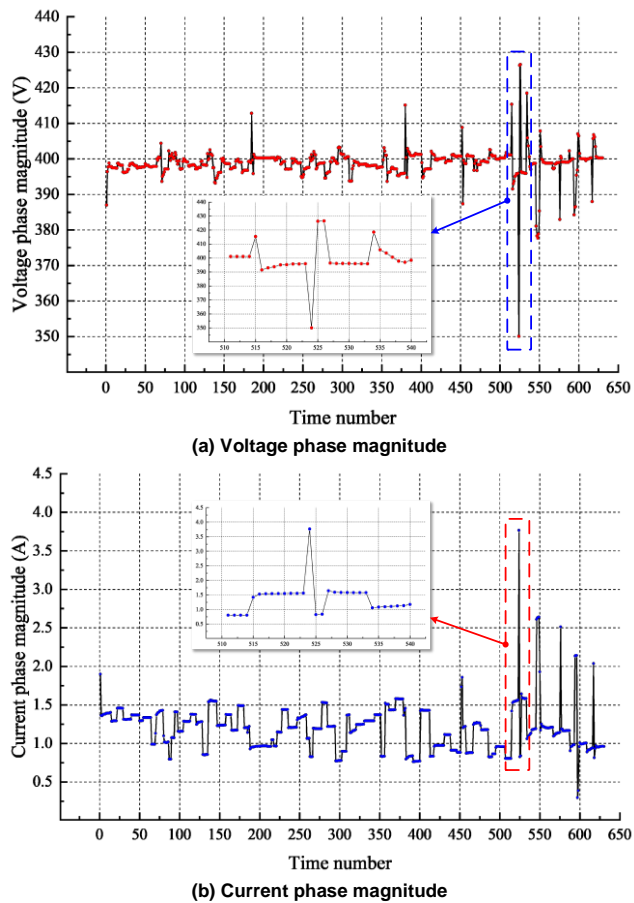

**FIGURE 10.** Impact of a remote trip command injection attack on the voltage and current phase magnitudes.

In summary, the established dynamic state transition model fully considers the closed-loop data flow characteristics of the microgrid CPS, is more in line with the constraint conditions of the spatiotemporal information and energy flow conversion of the microgrid CPS (perception-communication-decision-control), and can accurately describe the spatiotemporal deduction process of the data flow of FDIAs.

## VI. CONCLUSION AND FUTURE WORK

In this paper, the spatiotemporal conversion mechanism of the information flow and energy flow of the microgrid CPS was considered as the breakthrough point. A microgrid CPS system architecture with multi-agents as the core was constructed, and the discrete-continuous correlation and terminal structural characteristics of the CPS were characterized based on heterogeneous multi-groups. Considering the constraints of the temporal conversion of information flow and energy flow, a microgrid CPS coupling model driven by spatiotemporal events was established. A simulation verified that the proposed model can accurately describe the spatiotemporal deduction process of the FDIA data flow in the microgrid CPS, laying a theoretical foundation for supporting the safe and stable operation of microgrid CPS.

In subsequent research, the new generation of artificial intelligence technology (deep learning, reinforcement learning, etc.) will be introduced to consider the coupling characteristics and grid constraints of the power cyber-physical-social system (CPSS). Improving the identification of the power CPSS network cooperative attack antigen, adaptive defense, and other capabilities will be the focus of future research.

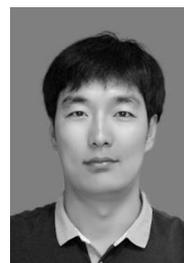

**XIAOYONG BO** (Student Member, IEEE) received his M.S. degree in Computer Application Technology in 2014 from Northeast Electric Power University, Jilin, China, where he is currently pursuing his Ph.D. degree in Electrical Engineering. His research interests include power cyber-physical systems modeling, attack identification, and defense in smart grid.

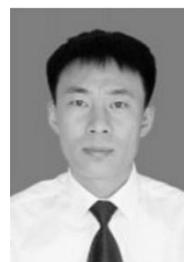

**XIAOYU CHEN** received his B.S. degree in Electrical Engineering and Automation from Changchun University. He is currently a Senior Engineer with the State Grid Jilin Electric Power Co., Ltd. Baishan Power Supply Company. His research interest is automated operation and maintenance of smart grid.

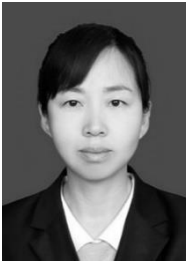

**HUASHUN LI** received her M.S. degree from Northeast Electric Power University. She is currently a Senior Engineer with the State Grid Jilin Electric Power Co., Ltd. Jilin Power Supply Company. Her research interests include automated operation and overhaul in smart grid.

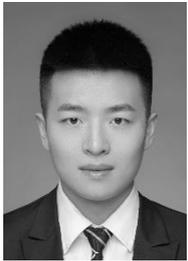

**YUNCHANG DONG** received his M.S. degree in 2019 from China Northeast Electric Power University, where he is currently pursuing his Ph.D. degree in Electrical Engineering. His research interests include power cyber-physical systems and information processing in smart grid.

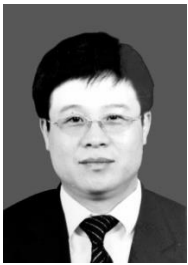

**ZHAOYANG QU** (Member, IEEE) received his Ph.D. degree in Electrical Engineering from China Northeast Electric Power University in 2010, and his M.S. degree from the Dalian University of Technology in 1988. He is currently a Professor and Doctoral Tutor with the School of Computer Science, Northeast Electric Power University. He is also Vice President of the Jilin Province Image and Graphics Society, Head of the Jilin Engineering Technology Research Center of Intelligent Electric Power Big Data Processing, and a Jilin Governor Baishan Scholar. His interests include network technology, smart grid, power information processing, and virtual reality. He has published more than 46 articles in SCI/EI international conference proceedings and journals. He is a member of the China Electric Engineering Society Power Information Committee. He has received the First Prize of the Jilin Province Science and Technology Progress Award. He is also a top-notch innovative talent in Jilin Province, and is designated as a young and middle-aged professional and technical talent with outstanding contributions. He has also presided over the completion of two national natural science funds.

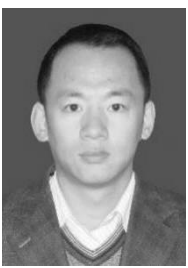

**LEI WANG** (Member, IEEE) is currently pursuing his Ph.D. degree in Electrical Engineering at Northeast Electric Power University, where he is an Associate Professor with the School of Information Engineering. His research interests include cyber-physical systems attacks and identification in smart grid. He has received the Second Prize of the Jilin Province Science and Technology Progress Award.

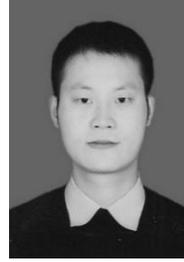

**YANG LI** (Senior Member, IEEE) received his Ph.D. degree in Electrical Engineering from North China Electric Power University (NCEPU), Beijing, China, in 2014. He is an Associate Professor with the School of Electrical Engineering, Northeast Electric Power University, Jilin, China. He is also a Postdoctoral Researcher with the Argonne National Laboratory, Lemont, USA, funded by the China Scholarship Council (CSC). His research interests include power system stability and control, integrated energy systems, renewable energy integration, and smart grid. He is also an Associate Editor of *IEEE Access*, *IET Renewable Power Generation*, and *Electrical Engineering*.